\documentstyle[12pt,epsf]{article}
\begin{document}
\title{
Large-$q$ expansion for the second moment correlation length 
in the two-dimensional $q$-state Potts model 
}
\author{
H. Arisue\\ 
Osaka Prefectural College of Technology \\ 
Saiwai-cho, Neyagawa, Osaka 572-8572, Japan
}
\date{January 2000}
\maketitle
\begin{abstract}
We calculate the large-$q$ expansion of the second moment correlation
length at the first order phase transition point
of the $q$-state Potts model in two dimensions 
both in the ordered and disordered phases to order 21 in $1/\sqrt{q}$. 
They coincide with each other to the third term of the series
but differ a little in higher orders. 
Numerically the ratio of the second moment 
correlation length in the two phases is not far from unity in all region
of $q>4$. The ratio of the second moment correlation length to the 
standard correlation length in the disordered phase is far from unity, 
which suggests that the second largest and smaller eigenvalues 
of the transfer matrix form a continuum spectrum
not only in the large-$q$ region but also in all the region of $q>4$. 
\end{abstract}
\newpage
\section{Introdunction}
The $q$-state Potts model\cite{Potts,Wu}
in two dimensions exhibits the first order phase transition for $q>4$. 
Many quantities\cite{Baxter1973} was solved exactly 
at the phase transition point, 
among which the correlation length $\xi^{(d)}$ in the disordered phase
\cite{Klumper,Buffenoir,Borgs} is included.
On the other hand, 
no analytic result had been known for the correlation length $\xi^{(o)}$ 
in the ordered phase at the phase transition point. 
Here the correlation length means the standard one 
defined from the ratio of the largest to the second largest eigenvalues 
of the transfer matrix and it is often called the true correlation length 
or the exponential correlation length.
Janke and Kappler\cite{Janke1994} examined the decay rate of the 
correlation function by the Monte Carlo simulation 
with large sizes of the lattice 
for $q=10,15$ and $20$ giving a result that was consistent 
with $\xi^{(o)} = \xi^{(d)}$ and they made a conjecture that this relation 
would be exact. 
Igl\'oi and Carlon\cite{Igloi} evaluated the largest and the second largest 
eigenvalues of the transfer matrix for $q=9$
by the density matrix renormalization group technique 
with a result that supports this conjecture.
Recently the author\cite{Arisue1999c} investigated analytically
the large-$q$ behavior of the eigenvalues of the transfer matrix.
He found that from the second largest to the $L$-th largest eigenvalues 
with $L$ the size of the lattice make a continuum spectrum 
for the thermodynamic limit both in 
the ordered and disordered phases at least in the large-$q$ region
and that $\xi^{(o)}=\xi^{(d)}$ at least to order $z^{3/2}$ 
(i.e., in the first 4 terms of the large-$q$ expansion).

Here we calculate the large-$q$ expansion of the second moment 
correlation length $\xi_{2nd}$ at the first order phase transition point
both in the ordered and disordered phases of this model.
The second moment correlation length also gives important 
informations on the spectrum of the eigenvalues of the transfer matrix.
The large-$q$ expansion of the Potts model in two dimensions 
was calculated to order $z^{10}$ ($z\equiv 1/\sqrt{q}$)
for the energy cumulants including the specific heat 
by Bhattacharya, Lacaze and Morel\cite{Bhattacharya1997}
using a graphical method.
Tabata and the author\cite{Arisue1999a,Arisue1999b} extended 
the large-$q$ series to order $z^{23}$ using the finite lattice 
method\cite{Enting1977,Creutz,Arisue1984}.
They also calculated the large-$q$ series for the magnetization 
cumulants including the magnetic susceptibility to order $z^{21}$. 
These long series enabled them to present the estimate for the numerical
values of the cumulants which is orders of magnitude more precise than that of
Monte Carlo simulations. The series also served to convince us that the
correctness of the conjecture by Bhattacharya, Lacaze and Morel 
on the behavior of the divergent quantities in the limit of $q\to 4$.
It is rather straightforward to apply the finite lattice method to 
the large-$q$ expansion of the second moment correlation length, since 
the algorithm is quite parallel with that of the finite lattice method 
for the low temperature expansion of the second moment correlation 
length for the Ising model in three dimensions\cite{Arisue1995}.
The situation is in contrast to the case of the standard 
correlation length where the continuum spectrum of the eigenvalues of 
the transfer matrix prevented us 
from applying the method used to obtain the low temperature series
for the Ising model in three dimensions\cite{Arisue1987, Arisue1994}.
The results of the large-$q$ expansion for the 
second moment correlation length was reported breafly 
in reference\cite{Arisue1999c} 
and here we describe them in detail.
\section{ Expansion series by the finite lattice method}
The model is defined on the $L_x \times L_y$ rectangular lattice 
by the partition function
\begin{equation}
 Z=\sum_{\{s_i\}} \exp{\left\{\beta \sum_{\langle i,j \rangle}\delta_{s_i,s_j}
          +\sum_{i}( h + \gamma_1 x_i + \gamma_2 y_i 
                          + \eta \mbox{\boldmath$r$}_i^2 )\delta_{s_i,1}
          \right\}}\;,  \label{eq:partitionf}
\end{equation}
where the spin variable $s_i$ at each site $i$ 
takes the values $1,2,\cdots,q$,
$\langle i,j \rangle$ represents the pair of nearest neighbor sites
and $\mbox{\boldmath$r$}_i=(x_i,y_i)$ is the coordinate of the site $i$.
The phase transition point $\beta_t$ for $h=\gamma_1=\gamma_2=\eta=0$ is given
by $\exp{(\beta_t)-1=\sqrt{q}}$. 
The fixed boundary condition should be taken for the ordered phase
in which all the spins outside the $L_x \times L_y$ lattice are fixed 
to be $\{ s_i=+1 \}$, and the free boundary condition should be taken 
for the disordered phase. 

 The second moment correlation length squared is defined by
\begin{equation}
      \xi_{2nd}^2 = \frac{\mu_2}{2 d \mu_0},
\end{equation}
where 
$\mu_2$ is the second moment of the correlation function 
\begin{equation}
   \mu_2 = 
       \lim_{L_x, L_y \rightarrow \infty}   (L_x L_y)^{-1} 
  \sum_{i,j} ( \mbox{\boldmath$r$}_i - \mbox{\boldmath$r$}_j )^2 
       \langle \delta_{s_i,1} \delta_{s_j,1} \rangle_c ,
\end{equation}
$\mu_0$ is the zeroth moment of the correlation function, 
i.e., the magnetic susceptibility, 
and $d(=2)$ is the dimensionality of the lattice. 
The second moment $\mu_2$ can be obtained by the derivative 
of the free energy density as
\begin{eqnarray}
   \mu_2 &=&   
      \lim_{L_x, L_y \rightarrow \infty}   (L_x L_y)^{-1} \nonumber\\&&
   \times\left. 
    2\left( \frac{\partial^2}{\partial h \partial \eta } 
        -  \frac{\partial^2}{\partial \gamma_1^2 } 
        -  \frac{\partial^2}{\partial \gamma_2^2 } 
    \right) 
             \ln{ Z(\beta,h,\eta,\gamma_1,\gamma_2) }
                 \right|_{h=\eta=\gamma_1=\gamma_2=0} .
\end{eqnarray}

 The algorithm of the finite lattice method 
to generate the large-$q$ expansion series 
for the second moment $\mu_2$ is the following. 
We define $H(l_x,l_y)$ for each $l_x\times l_y$ lattice
($1\le l_x\le L_x, 1\le l_y\le L_y$) as
\begin{equation}
   \!\!\!\!\!\!\!\!\!\!\!\!\!
   H(l_x,l_y) =   \left.
      2 \left( \frac{\partial^2}{\partial h \partial \eta } 
        -  \frac{\partial^2}{\partial \gamma_1^2 } 
        -  \frac{\partial^2}{\partial \gamma_2^2 }  \right) 
          \ln{ Z(l_x,l_y)} 
      \right|_{h=\eta=\gamma_1=\gamma_2=0 },
\end{equation}
where $Z(l_x,l_y)$ is the partition function for the $l_x\times l_y$ lattice
with the fixed and free boundary condition 
for the ordered and disordered phase, respectively, 
and define $W$ of each lattice recursively as
\begin{eqnarray}
W(l_x,l_y) &=& H(l_x,l_y) \nonumber\\
&& - \sum_{\scriptstyle 1\le l_x^{\prime}\le l_x, 1\le l_y^{\prime}\le l_y
\atop\scriptstyle  (l_x^{\prime},l_y^{\prime})\ne (l_x,l_y)}
(l_x-l_x^{\prime}+1)(l_y-l_y^{\prime}+1)W(l_x^{\prime},l_y^{\prime})
\;. \label{eq:W}
\end{eqnarray}
 Note that $H(l_x,l_y)$ and $W(l_x,l_y)$ depend 
on the size $l_x$ and $l_y$ but not on the position of the origin of the 
coordinate. 
 Then the second moment of the correlation function in the thermodynamic 
limit is given by
\begin{equation}
   \mu_2 =  \sum_{l_x,l_y}  W( l_x,l_y ) .  \label{eqn:B}
\end{equation}

 We can prove\cite{Arisue1984} that the Taylor expansion 
of $W(l_x,l_y)$ with respect to $z\equiv 1/\sqrt{q}$
includes the contribution from all the clusters of polymers 
in the standard cluster expansion\cite{Muenster}
that can be embedded into the $l_x\times l_y$ lattice 
but cannot be embedded into any of its rectangular 
$l_x^{\prime}\times l_y^{\prime}$ 
sub-lattices with 
$ l_x^{\prime}\le l_x,l_y^{\prime}\le l_y$. 
 It is straightforward to understand from the discussion 
for the large-$q$ expansion of the magnetic susceptibility 
in reference\cite{Arisue1999b}
that the series expansion of $W(l_x,l_y)$ 
starts from the order of $z^n$ with 
$n=l_x+l_y$ in the ordered phase and 
$n=l_x+l_y-2$ in the disordered phase, respectively. 
 So in order to obtain the expansion series to order $z^N$, 
we should take all the finite-size rectangular lattices 
for the summation in equation~(\ref{eqn:B}) that satisfy 
$l_x+l_y\le N$ in the ordered phase and 
$l_x+l_y-2\le N$ in the disordered phase, respectively.

Using these algorithm of the finite lattice method we have calculated 
the series for the second moment $\mu_2$ at the first order phase 
transition point to order $N=21$ in $z$ 
both in the ordered and disordered phases as 
\begin{equation}
     \mu_2 = \sum_{n} a_n z^n.
\end{equation}
The coefficients of the series are listed in table 1.
We have checked that all of $W(l_x,l_y)$
with $l_x+l_y-2\le N$ for the ordered phase 
and with $l_x+l_y\le N$ for the disordered phase 
have the correct order in $z$ as described above.
\begin{table}[tb]
\caption{
The large-$q$ expansion coefficients $a_n$
for the second moment $\mu_2$ of the correlation function
at $\beta=\beta_t$ in the ordered and disordered phases.
         }
\begin{tabular}{rrr}
\hline
$n$      & $a_n$(ordered)   & $a_n$(disordered) \\
\hline
$ 0$ &$             0$ &$             0$ \\
$ 1$ &$             0$ &$             0$ \\
$ 2$ &$             0$ &$             0$ \\
$ 3$ &$             4$ &$             4$ \\
$ 4$ &$            44$ &$            44$ \\
$ 5$ &$           344$ &$           336$ \\
$ 6$ &$          2140$ &$          2064$ \\
$ 7$ &$         11676$ &$         11164$ \\
$ 8$ &$         57440$ &$         54652$ \\
$ 9$ &$        261884$ &$        248456$ \\
$10$ &$       1121092$ &$       1062320$ \\
$11$ &$       4560648$ &$       4320948$ \\
$12$ &$      17761800$ &$      16840044$ \\
$13$ &$      66657396$ &$      63279396$ \\
$14$ &$     242180712$ &$     230307200$ \\
$15$ &$     855255360$ &$     815001104$ \\
$16$ &$    2945011236$ &$    2812875600$ \\
$17$ &$    9914462596$ &$    9493154576$ \\
$18$ &$   32704960680$ &$   31397031360$ \\
$19$ &$  105910040516$ &$  101949411404$ \\
$20$ &$  337252952816$ &$  325538033480$ \\
$21$ &$ 1057475140292$ &$ 1023600956088$ \\
\hline
\end{tabular}
\end{table}

Combining with the large-$q$ series of the magnetic susceptibility
given in reference\cite{Arisue1999b}, we obtain the series 
for the second moment correlation length squared $\xi_{2nd}^2$ as
\begin{equation}
     \xi_{2nd}^2 = \sum_{n} b_n z^n,
\end{equation}
which are listed in table 2. 
The obtained expansion coefficients for the ordered and disordered phases
coincide with each other to order $z^{3}$ 
and differ a bit from  each other in higher orders.
\begin{table}[tb]
\caption{
The large-$q$ expansion coefficients $b_n$
for the second-moment correlation length squared $\xi_{2nd}^2$
at $\beta=\beta_t$ in the ordered and disordered phases.
         }
\begin{tabular}{rrr}
\hline
$n$      & $b_n$(ordered)   & $b_n$(disordered) \\
\hline
$ 0$ &$          0$ &$          0$ \\
$ 1$ &$          1$ &$          1$ \\
$ 2$ &$          7$ &$          7$ \\
$ 3$ &$         35$ &$         35$ \\
$ 4$ &$        154$ &$        157$ \\
$ 5$ &$        611$ &$        635$ \\
$ 6$ &$       2237$ &$       2363$ \\
$ 7$ &$       7720$ &$       8271$ \\
$ 8$ &$      25461$ &$      27569$ \\
$ 9$ &$      80852$ &$      88271$ \\
$10$ &$     249212$ &$     273775$ \\
$11$ &$     748239$ &$     825956$ \\
$12$ &$    2198991$ &$    2436937$ \\
$13$ &$    6335396$ &$    7044039$ \\
$14$ &$   17961597$ &$   20027792$ \\
$15$ &$   50129172$ &$   56038776$ \\
$16$ &$  138121342$ &$  154746684$ \\
$17$ &$  375778895$ &$  421876179$ \\
$18$ &$ 1011869143$ &$ 1137902174$ \\
$19$ &$ 2695052762$ &$ 3035669883$ \\
\hline
\end{tabular}
\end{table}

\section{Analysis of the series}

It is known\cite{Caselle1999} that in the limit of the large 
correlation length
\begin{equation}
   \frac{\xi_{2nd}^2}{\xi_1^2}
     \to \frac{\sum_{j=1}^{\infty} c_j^2 (\xi_j/\xi_1)^3}
               {\sum_{j=1}^{\infty} c_j^2 (\xi_j/\xi_1)} 
         \label{eq:xisecasym}
\end{equation}
with $\xi_j\equiv -\ln{(\Lambda_j/\Lambda_0)}$
and $c_j=<\Lambda_0|{\cal O}|\Lambda_j>$
where $\Lambda_j(j=0,1,2,\cdots)$ are the eigenvalues of the
transfer matrix with $\Lambda_0$ the largest one, 
$|\Lambda_j>$ is the eigenstate of the transfer matrix 
corresponding to the eigenvalue $\Lambda_j$
and ${\cal O}=\sum_i \delta_{s_i,1}$ with the summation for $i$ running 
over the sites with a fixed coordinate of $y_i$.
The $\xi_1\equiv -\ln{(\Lambda_1/\Lambda_0)}$ is the standard correlation 
length.
Later on we will call the eigenstate $\Lambda_1$ the first excited state
and $\Lambda_j$($i=2,3,\cdots$) the higher excited states.  

The standard correlation length $\xi_{1}^{(d)}$ at the phase transition point 
in the disordered phase is known exactly 
with its asymptotic behavior for $q \to 4$ as 
\begin{equation}
   \xi_{1}^{(d)} \sim  \frac{1}{8\sqrt{2}}
     \exp{\left(\frac{\pi^2}{2\theta} \right)},
\end{equation}
where $2\cosh{\theta} = \sqrt{q}$ 
($\theta \sim \sqrt{q-4}/2$ for $q\to 4$). 
It is quite natural to expect that all of the $\xi_j$'s behave like
\begin{equation}
 \xi_{j}^{(o,d)} \sim  C_j^{(o,d)} \exp{\left(\frac{\pi^2}{2\theta} \right)}
\end{equation}
for $q \to 4$ both in the ordered and disordered phases,
in which case 
\begin{equation}
   \xi_{2nd}^{(o,d){}^2} \sim 
        A^{(o,d)} \exp{\left(\frac{\pi^2}{\theta} \right)}.
            \label{eq:asymp_form}
\end{equation}

To check the validity of this conjecture we follow the method 
in reference\cite{Arisue1999a,Arisue1999b}
adopted to analyze the large-$q$ series of the energy and 
magnetization cumulants
including the specific heat and the magnetic susceptibility.
The method was used to convince the validity of the Bhattacharya-Lacaze-Morel
conjecture on the asymptotic behavior of these quantities for $q\to 4$.
The latent heat ${\cal L}$ is known exactly with the asymptotic form 
for $q\to 4$ as\cite{Bhattacharya1994} 
\begin{equation}
{\cal L} \sim 3\pi \exp{\left(-\frac{\pi^2}{4\theta} \right)}\;, 
        \label{eq:latentheat}
\end{equation}
so, if $\xi_{2nd}^{(o,d)}{}^2$ has the asymptotic form 
in equation~(\ref{eq:asymp_form}), 
the product $\xi_{2nd}^2{\cal L}^{p}$ is a smooth function of $z$
for $q\to 4$ when $p=4$, and the Pad\'e approximants of the large-$q$ 
series for this product are expected to converge for $p=4$.
The results are given in figure 1. 
We can see that the convergence is really good around $p=4$ 
both in the ordered and disordered phases. 
\begin{figure}[tb]
\epsfxsize=13cm
\epsffile{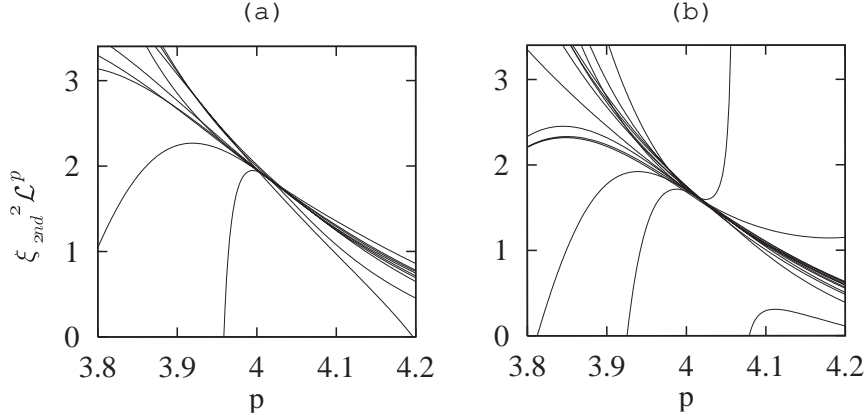}
\caption{
The Pad\'e approximants of the second moment correlation length squared 
multiplied by the $p$-th power of the latent heat
(a) in the disordered phase and (b) in the ordered phase.
         }
\end{figure}

The estimated values of the second moment correlation length 
by setting $p=4$ are listed in table.3. 
We note that the ratio 
of the estimated values of 
the second moment correlation length 
in the ordered to disordered phases 
$\xi_{2nd}(\mbox{ordered})/\xi_{2nd}(\mbox{disordered})$
is not far from unity.
The ratio is 0.935(5) even for $q\to 4$.
\begin{table}[tb]
\caption{Estimates of the second moment correlation length $\xi_{2nd}$ 
for each $q$ in the ordered and disordered phases.}
\begin{tabular}{rll}
\hline
    $q$  & $\xi_{2nd}$(ordered)  &  $\xi_{2nd}$(disordered) \\
\hline
 5 & 1261(4)      & 1349(6)     \\
 6 & 83.1(1)      & 88.5(1)     \\
 7 & 25.89(1)     & 27.45(2)    \\
 8 & 13.140(2)    & 13.855(5)   \\
 9 & 8.3440(5)    & 8.751(4)    \\
10 & 5.9965(2)    & 6.260(1)    \\
12 & 3.79944(3)   & 3.9355(1)   \\
15 & 2.480927(4)  & 2.54815(1)  \\
20 & 1.6376632(4) & 1.667680(1) \\
30 & 1.0633984(1) & 1.0742339(1)\\
\hline
\end{tabular}
\end{table}
\begin{figure}[tb]
\epsfxsize=8cm
\epsffile{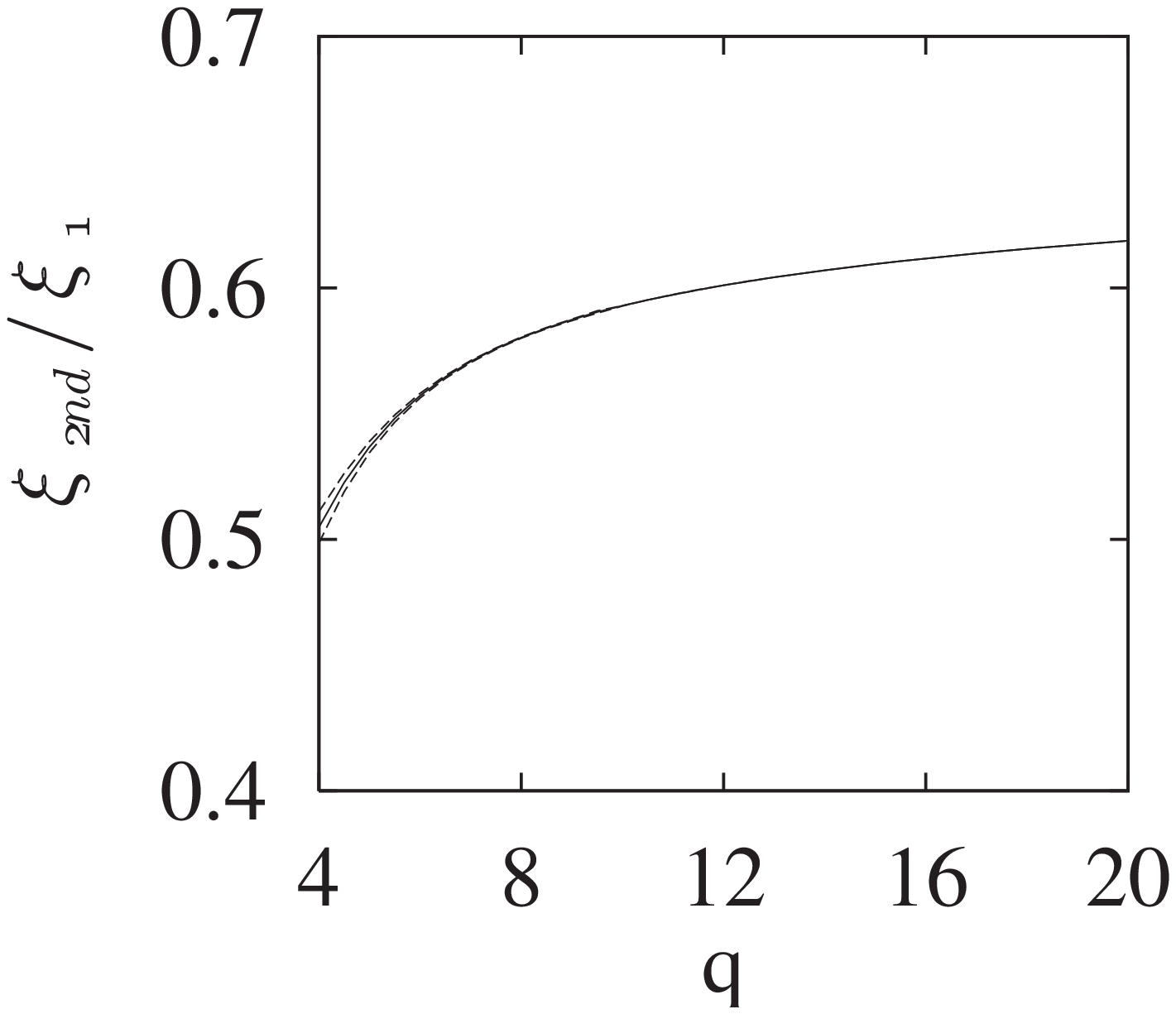}
\caption{
The ratio $\xi_{2nd}/\xi_1$ of the second moment correlation length
to the standard correlation length
in the disordered phase.
         }
\end{figure}

Another interesting and important quantity is the ratio of the 
second moment correlation length $\xi_{2nd}$ to the standard 
correlation length $\xi_1$. 
From equation~(\ref{eq:xisecasym}) 
we know that the ratio $\xi_{2nd}/\xi_1$ should be less than 
unity in the limit of the large correlation length.
If the higher excited states ($i=2,3,\cdots$)
did not contribute so much, 
this ratio would be close to unity,
as is the case in the Ising model on the simple cubic lattice,
where $\xi_{2nd}/\xi_1=0.970(5)$ 
at the critical point in the disordered phase
\cite{Caselle1999,Campostrini1998}.

In figure 2 we plot the ratio $\xi_{2nd}/\xi_1$ for the Potts model
in the disordered phase. We use the value of $\xi_{2nd}$
estimated above from the Pad\'e analysis of $\xi_{2nd}^2{\cal L}^4$ 
and the exact value of $\xi_1$. 
The ratio is much less than unity in the region of $q$ 
where the correlation length is large enough. 
It approaches $0.505(6)$ for $q \to 4$. 
This result implies that the contribution of 
the higher excited states 
is important in the Potts model in two dimensions.
It is consistent with the fact that the eigenvalue of the 
transfer matrix for the first excited state locates 
at the edge of the continuum spectrum of the eigenvalues of the transfer
matrix at least in the large-$q$ region
and it strongly suggests that 
these excited states ($i=1,2,\cdots,L_x$) form a continuum spectrum 
not only in the large-$q$ region but also even in the limit of $q\to 4$.

\section{Summary}
We calculated the large-$q$ expansion of the second moment correlation
length in the ordered and disordered phases of the $q$-state Potts model
in two dimensions
and found that they coincide with each other to the third term of the 
expansion but differ a little from each other in higher orders. 
This suggests that, although the first few terms of the large-$q$ expansion
for the standard correlation length really coincide 
in the ordered and disordered phases, 
their higher order terms might be different from each other. 
Note, however, that it is not conclusive,
since the second moment correlation length depends not only on the spectrum
of the eigenvalues of the transfer matrix 
but also on the overlapping amplitude $c_i$ which appeared 
in equation~(\ref{eq:xisecasym}).
Numerically the ratio $\xi_{2nd}^{(o)}/\xi_{2nd}^{(d)}$ is not far from unity 
even in the limit of $q\to 4$.
We also found that $\xi_{2nd}^{(d)}/\xi_1^{(d)}$ is far from unity 
for all region of $q>4$. 
It implies that higher excited states
give significant contributions as well as the first excited state
and it strongly suggests that 
these excited states form a continuum spectrum 
(i.e., there is  no particle state in the language of the field theory)
not only in the large-$q$ region but also in all the region of $q>4$.
Finally we point out that it is worthwhile to reanalyze 
the previous Monte Carlo data for the correlation function 
on the assumption that this would be the case.

\section*{Acknowledgments}
The author would like to thank W. Janke, A. Sokal and K. Tabata 
for valuable discussions.


\end{document}